\documentclass[conference]{IEEEtran}

\usepackage{cite}
\usepackage{amsmath,amssymb,amsfonts}
\usepackage{graphicx}
\usepackage{textcomp}
\usepackage{enumitem}

\usepackage{amsmath}
\usepackage{amssymb}
\usepackage{amsthm}
\usepackage{amsfonts}
\usepackage{epsfig}
\usepackage{bm}
\usepackage{graphicx}
\usepackage[font={small}]{caption}
\usepackage{subcaption}
\usepackage{float}
\usepackage{algorithm}
\usepackage{algorithmicx}
\usepackage{algpseudocode}
\usepackage{soul}
\usepackage{xcolor}
\usepackage{comment}
\usepackage{cite}
\usepackage{bbold}

\newtheorem{thm}{Theorem}
\newtheorem{assumption}{Assumption}
\newtheorem{rem}{Remark}
\newtheorem{lem}{Lemma}
\newtheorem{prop}{Proposition}
\newtheorem{cor}{Corollary}
\newtheorem{defn}{Definition}
\newtheorem{exmp}{Example}

\newcommand{\R}{\mathbb{R}}

\newcommand{\K}{\mathcal{K}}

\newcommand{\A}{\mathcal{A}}

\newcommand{\B}[1]{\boldsymbol{#1}}

\newcommand{\be}{\begin{equation}}
\newcommand{\ee}{\end{equation}}
\newcommand{\ba}{\begin{aligned}}
\newcommand{\ea}{\end{aligned}}

\newcommand{\lba}{\left[ \begin{array}}
\newcommand{\ear}{\end{array} \right]}

\newcommand{\one}{\mathbb{1}}
\newcommand{\zero}{\mathbf{0}}

\newcommand{\emily}[1]{\textcolor{blue}{{#1}}}

\newcommand{\walden}[1]{\textcolor{orange}{{#1}}}

\IEEEoverridecommandlockouts

\begin{document}
\title{A Convex Parameterization of Controllers\\ Constrained to use only  Relative Measurements
}
\author{Walden Marshall, Bassam Bamieh, and Emily Jensen
\thanks{Walden Marshall and Emily Jensen are with the Electrical, Computer \& Energy Engineering Department at the University of Colorado, Boulder (e-mails: $\{$walden.marshall, ejensen$\}$ @colorado.edu).}
\thanks{ Bassam Bamieh is with the Mechanical Engineering Department at the University of California, Santa Barbara (email: bamieh@ucsb.edu).}}

\maketitle



\begin{abstract} The optimal controller design problem for systems equipped with sensors that measure only relative, rather than absolute, quantities is considered. This relative measurement structure is formulated as a design constraint; it is demonstrated that the resulting constrained controller design problem can be written as a convex program. Certain additional network structural constraints can be incorporated into this formulation, making it especially useful in distributed or networked settings. An illustrative example highlights the advantage of the proposed methodology over the standard formulation of the output feedback controller design problem. A numerical example is provided.
\end{abstract}


\section{Introduction}

{Recent technological advances enable increasingly distributed sensing and control in engineering systems. Examples include distributed systems, e.g., networks of individual subsystems comprising a satellite constellation or vehicle platoon, or distributed sensing and actuation in a single flexible system, e.g., the body of a single soft robot \cite{wang2018toward, correll2014soft, kim2013soft}  or for human sensory feedback \cite{park2023skin}. A vast area of research has emerged to tackle some of the resulting challenges, much of which has focused on the constraints resulting from the limited communication or interaction between spatially distributed controller units, which lead to delay or sparsity constraints on the underlying control policy \cite{dullerud2004distributed, nayyar2013decentralized,mahajan2012information}.

Another limitation may arise from the fundamental properties of sensors or actuators available to these systems. Indeed, a large number of systems with distributed sensing can measure only relative (rather than absolute) distances or velocities between subsystems or regions, and/or can only actuate subsystems or regions relative to each other. For example, a satellite constellation may be equipped only with sensors measuring relative position and velocities between neighboring satellites, and/or actuation in the system may be limited to relative forces and torques between satellites through superconducting electromagnetic coils \cite{ahsun2006dynamics,abbasi2022decentralized}. Soft robotic systems may sense bending or torsional motion through the relative position differences between segments of the body using, e.g., optical strain sensors \cite{wang2018toward}. 

It is well-known that delay/sparsity constraints on the controller due to the limited communication or interaction between spatially distributed units of a system make the optimal controller design problem significantly more challenging than its centralized counterpart. In general, this constrained controller design problem is non-convex without known tractable methods for solving in general settings. A well-known exception is the case of quadratic invariance \cite{rotkowitz2005characterization}.

It is not surprising that access to only relative sensors or actuators may introduce additional challenges for controller design. Recent work \cite{jensen2020gap} has highlighted that the combination of these two distinct properties - relative measurement structure and sparse network interaction - may be particularly problematic; one optimal controller design problem subject to network structural constraints utilizing the System Level Synthesis framework \cite{wang2019system} was shown to become infeasible when measurements were restricted to be relative.  (rather than absolute).
}

In this work, 
\emph {we provide a characterization of relative measurement structure as an explicit design constraint} (Theorem~\ref{prop:recover_original_K}).  
Although  network control structures are often encoded as constraints on the controller design problem, an analogous characterization of {relative measurement structures} as a design constraint in general settings appears to be lacking. Such a constraint has been utilized in the case that the relative difference between \emph{all} possible combinations of states can be inferred \cite{jensen2020gap, oral2019disorder, tegling2017coherence}, but our result holds more generally. This is important in networked settings, where the sensing or communication graph among subsystems may not be connected.
We show that this design constraint allows the \emph{relative measurement controller design problem to be formulated as a convex program }(Theorem~\ref{main_result}). 
 Moreover, we utilize the framework of quadratic invariance \cite{rotkowitz2005characterization} to demonstrate that \emph{certain network structural constraints can be incorporated into this formulation while preserving convexity}. Thus, this work takes a step toward developing a computationally tractable formulation of the optimal controller design problem for systems with sensors that measure only relative quantities.

The rest of this paper is structured as follows: Section~\ref{sec:setup} presents the optimal controller design problem and introduces relative measurements and network structure. In Section~\ref{sec:counterexample}, an example  illustrates that standard output feedback control design techniques may not be equipped to handle the simultaneous presence of relative measurement and network structure requirements, motivating an alternative approach.
{In Section~\ref{sec:define_relative}, relative measurement structure is formulated as a design constraint; this allows the optimal controller design problem for relative measurement systems to be written as a convex program
in Section~\ref{sec:convex}. In Section~\ref{sec:incorporate_control_structure}, certain network structural constraints are incorporated in this formulation while preserving convexity; a numerical example is provided.}

\color{black}
\section{Problem Set Up} \label{sec:setup}
        
\subsection{Notation \& Preliminaries}
We begin by introducing the optimal controller design problem and formalizing relative measurement and network structure.
We use bold face lowercase lettering to denote a signal in the frequency domain, and bold face upper case lettering for a linear time-invariant (LTI) system or its corresponding transfer matrix. For simplicity, we present our results in the continuous-time setting; the discrete-time setting follows analogously.

We consider a generalized plant ${\B P}$ with dynamics
    \be \ba \label{eq:P}
        \dot{ {x}}(t) &= A { x}(t) + B_1 { w}(t) + B_2 { u} (t)\\
        z(t) & = C_1 x(t) + D_{12} u(t) \\
        y(t) & = C_2 x(t)
    \ea \ee 
where $x(t) \in \R^n, w(t) \in \R^q, u(t) \in \R^l, z(t) \in \R^r$ and $y(t) \in \R^p$ are the state, exogenous disturbance, control signal, performance output and measurement, at time $t$, respectively. We omit the dependence on $t$ when clear from context.

When ${\B P}$ is in feedback with an LTI controller ${\B u} = {\B K}{\B y}$, the closed-loop map from ${\B w}$ to ${\B z}$ takes the form of a linear fractional transformation:
    $$
       {\B z} = \mathcal{F}({\B P}; {\B K}) {\B w} =  \left({\B P}_{zw} + {\B P}_{zu} {\B K} (I - {\B P}_{yu} {\B K})^{-1} { \B P}_{yw}\right) {\B w},
    $$
where the transfer matrix representation ${\B P}$ of \eqref{eq:P} is partitioned as, 
$
    \lba{c} {\B z} \\ {\B y} \ear = \lba{cc} {\B P}_{zw} & {\B P}_{zu} \\ {\B P}_{yw} & {\B P}_{yu} \ear\lba{c} {\B w} \\ {\B u} \ear.
$
The optimal controller design problem takes the form
\be \ba \label{eq:general}
         & \inf_{{\B K}} ~~ \| \mathcal{F}({\B P}; {\B K}) \|\\
         & ~{\rm s.t.}~~ {\B K} \in \mathcal{K}_{\rm stab.}({\B P}),
    \ea \ee 
where $\| \cdot\|$ is an operator norm and $\mathcal{K}_{\rm stab.}({\B P})$ is the set of controllers that internally stabilize ${\B P}$. $\mathcal{K}_{\rm stab.}({\B P})$ can be characterized as \cite{francis1987course},\cite{rotkowitz2005characterization}:
\be \label{eq:set_unstable_plant}
    \ba
        &\mathcal{K}_{\rm stab.}({\B P})  = \\
       & \left\{ {\B K};~ {\B K} \!= \!{\B K}_{\rm nom.} \! - \! h\left(h\left({\B K}_{\rm nom.}, {\B P}_{yu}\right),{\B Q}\right) , ~ {\B Q}~ \text{stable} \right\},
    \ea
\ee
where ${\B K}_{\rm nom.}$ is a nominal stabilizing controller, and following the notation of \cite{rotkowitz2005characterization},
\be 
    h({\B G}, {\B K}) = -{\B K}(I - {\B G} {\B K})^{-1}.
\ee 

With constraints on the controller, \eqref{eq:general} takes the form
    \be \ba  \label{eq:general_constrained}
         & \inf_{\B K} ~~ \| \mathcal{F}({\B P}, {\B K}) \|\\
         & ~{\rm s.t.}~~ {\B K} \in \mathcal{K}_{\rm stab.}({\B P}) \cap\mathcal{S}
    \ea \ee
For general $\mathcal{S}$, \eqref{eq:general_constrained} is nonconvex without known computationally tractable methods for solving. A notable exception is the setting of quadratic invariance. A subspace 
$\mathcal{S}$ is said to be \emph{quadratically invariant} (QI) under ${\B G}$ if $\B{K}\in\mathcal{S}\implies\B{K}\B{G}\B{K}\in\mathcal{S}$ \cite{rotkowitz2005characterization}.
It is known that if $\mathcal{S}$ is QI under 
 $\B{P}_{yu}$, 
then \eqref{eq:general_constrained} can then be rewritten as 
\be \ba  \label{eq:phiQ}
         & \inf_{\B Q} ~~ \| \phi({\B Q})\|\\
         & ~{\rm s.t.}~~ {\B Q} \in 
         \mathcal{S}, ~{\B Q}~\mathrm{stable}
    \ea \ee
where $\phi({\B Q})$ is affine in ${\B Q}.$ For instance, when ${\B P}_{yu}$ is stable, $\phi({\B Q})$ takes the form ${\B P}_{zw} + {\B P}_{zu} {\B Q} {\B P}_{yw}$ \cite{rotkowitz2005characterization}. 

\subsection{Relative Measurement Structure}
We assume the measurement $y(t)$ of system \eqref{eq:P} contains only \emph{relative} (rather than absolute) quantities. Specifically, we assume each entry of $y$ is the difference of two distinct states: 
\begin{assumption} \label{assumption:C2} 
Each row of $C_2$ contains exactly one entry of $1$ and one entry of $-1$. 
\end{assumption} 

\begin{exmp} \label{exmp:3components}
    The relative quantities contained in the measurement vector
\be \label{eq:exmp_rel_structure}
    y  = \lba{c} x_1 - x_3 \\ x_4 - x_2 \ear = {\lba{ccccc} 1 & 0 & -1 & 0 &0\\
    0 & -1 & 0 & 1 & 0  \ear}x{=:C_2} x
\ee
for a system with state $x(t) \in \mathbb{R}^5$
is illustrated in Figure~\ref{fig:three_components}.
\vspace{-3mm}
    \setlength{\belowcaptionskip}{-5pt}
\begin{figure}[h]
    \centering
    \includegraphics[width = .35\textwidth]{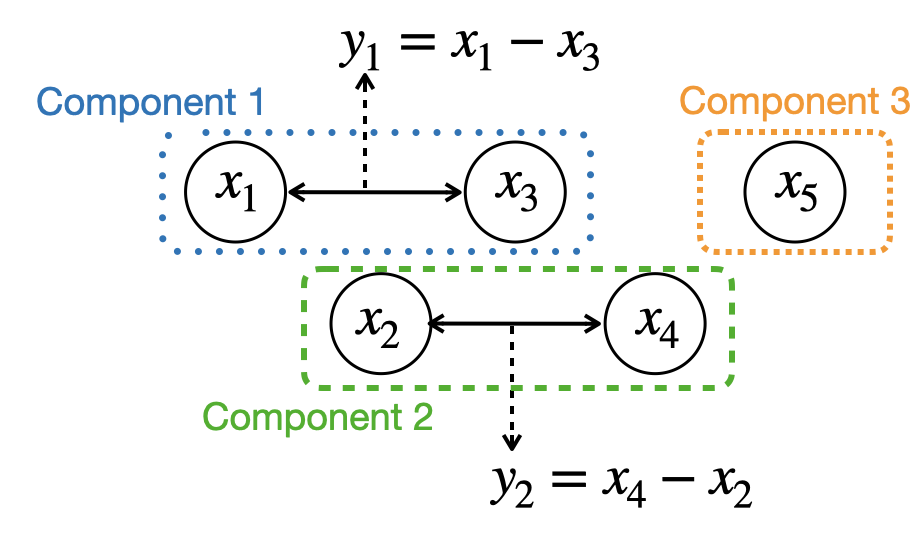}
    \caption{Graph corresponding to relative measurement structure of \eqref{eq:exmp_rel_structure}}
    \label{fig:three_components}
\end{figure}
This relative measurement structure \eqref{eq:exmp_rel_structure} partitions the set of states into three connected components. This structure is captured by a graph with adjacency matrix 
        \setlength{\arraycolsep}{4pt}
    \be 
        \mathcal{A}_{\rm meas.} (C_2) = \lba{ccccc} 0 & 0 & 1 & 0 & 0  \\ 0 & 0 & 0 & 1 & 0  \\ 1 &0 &0&0&0 \\ 0&1&0&0 & 0 \\ 0&0&0&0&0 \ear.
    \ee
\end{exmp}
To capture the relative measurement architecture more generally, let each state $x_i, ~i=1,...,n$ of system \eqref{eq:P} correspond to a node in a graph. Construct a graph with adjacency matrix $\mathcal{A}_{\rm meas.}(C_2)$ according to 
    \be \label{eq:Ameas}
        \mathcal{A}_{\rm meas.}(i,j) = \begin{cases}
            1, ~\exists~ \ell ~{\rm s.t.}~C_2(\ell,i) \ne 0~{\rm and}~C_2(\ell,j) \ne 0\\
            0, ~ {\rm else.}
        \end{cases}
    \ee 
In effect, there is an edge between two nodes in this graph if the relative difference between the corresponding two states is measured; this graph is well-defined under Assumption 1. We characterize the set of controllers that satisfy the relative measurement structure described by $\mathcal{A}_{\rm meas.}(C_2)$ in Section~\ref{sec:define_relative}.

\subsection{Network Control Structure} \label{sec:network_structure}

In the distributed setting, each subcontroller may have access to only a subset of system measurements and with limited or delayed communication between these subcontrollers. Corresponding delay and sparsity constraints (independent of the relative measurement structure) are characterized as a subspace constraint, ${\B K} \in \mathcal{S}_{\rm net.}$, on the controller.

In what follows, we consider the design problem \eqref{eq:general_constrained} in the case that the controller is limited by access to only relative measurements, and possibly also by delay/sparsity constraints introduced by a network control structure. 

\section{Motivating Example} \label{sec:counterexample}
We consider the controller design problem for
a system composed of four subsystems with dynamics of the form 
    \setlength{\arraycolsep}{1.8pt}
    \be \label{eq:exmp_dynamics}
        \lba{c} \dot{x}_1 \\ \dot{x}_2 \\ \dot{x}_3 \\ \dot{x}_4 \ear {\text =} \underbrace{\lba{cccc} a_{11} & a_{12} & a_{13} & a_{14} \\ 0 & a_{22} & a_{23} & a_{24} \\ 0 & 0& a_{33} & a_{34} \\ 0 &0&0&a_{44}\ear }_{=:A}\lba{c} x_1 \\ x_2 \\ x_3 \\ x_4 \ear + \lba{c} u_1 \\ u_2 \\ u_3 \\ u_4 \ear + \lba{c} w_1 \\ w_2 \\ w_3 \\ w_4 \ear ,
    \ee 
where each $a_{ii} <0$ so that $A$ is Hurwitz, $x_i$ is the internal state of subsystem $i$, and $u_i$ and $w_i$ are the control and disturbance at subsystem $i$, respectively. 

When the controller is restricted to access only relative measurements or when the controller is subject to network structure constraints, quadratic invariance allows us to convert the design problem to the convex form \eqref{eq:phiQ}. However, when both constraints are simultaneously imposed, quadratic invariance does not hold; this motivates our reformulation.

{\bf Relative measurement structure alone:} 
Sensors throughout the system measure $(x_i - x_j)$ for all $j >i$. The controller to be designed is then an output feedback controller ${\B u} = {\B K}{\B y}$ where 
 \be\label{eq:exmp_meas}
        y =\lba{c} y_1 \\ y_2 \\ y_3 \\ y_4 \\ y_5 \\ y_6  \ear 
        = \lba{cccc} 1 & -1 &0&0\\
        1& 0 & -1 & 0 \\
        1 & 0 & 0 & -1\\
        0 & 1 & -1 & 0 \\
        0 & 1 &0&-1 \\ 0 &0&1 &-1\ear\lba{c} x_1 \\ x_2 \\ x_3 \\ x_4 \ear {=:C_2} x.
    \ee 
The controller design problem is of the form \eqref{eq:general} and can be tractably solved with standard approaches\cite{youla1976modern}.
\color{black}

{\bf Network control structure alone:} 
The controller to be designed is spatially distributed with one subcontroller at each subsystem. We assume that sensors located at subsystem $i$ measure $x_j$ for all $j\geq i$ and that only the subcontroller at $i$ has access to these measurements, i.e. $u_i$ may only depend on $x_j$ for $j\geq i$. The control policy takes the form
\be \lba{c} {\B u}_1 \\ {\B u}_2 \\ {\B u}_3 \\ {\B u}_4 \ear {{=}} \lba{cccccc} {\B R}_{11} & {\B R}_{12} & {\B R}_{13} &{\B R}_{14}\\
         0& {\B R}_{22} & {\B R}_{23} & {\B R}_{24} \\
         0 & 0&  {\B R}_{33} & {\B R}_{34}\\
         0 &0&0  &{\B R}_{44}\ear \lba{c}{\B x}_1 \\ {\B x}_2 \\ {\B x}_3\\ {\B x}_4 \ear .
    \ee 
This upper triangular structure is QI with respect to the mapping ${\B P}_{xu} = (sI - A)^{-1}B_2$, so the (state feedback) controller optimization can be converted to the form \eqref{eq:phiQ} and solved with standard approaches \cite{rotkowitz2005characterization}.

{\bf Relative measurement and network control structure:} We now apply both requirements by assuming that sensors at subsystem $i$ measure the relative difference $(x_i-x_j)$ for $j>i$ and that only subcontroller $i$ will have access to this measurement,  i.e. the control signal $u_i$ may depend only on $(x_i-x_j)$ for $j>i$. A standard output feedback approach aims to design  a control law ${\B u} = {\B K}{\B y}$ that satisfies the subspace constraint ${\B K} \in \mathcal{S}$, where $\mathcal{S}$ encodes the sparsity structure:
\setlength{\arraycolsep}{1.5pt}
\be  \label{eq:output_structure}
         \lba{c} {\B u}_1 \\ {\B u}_2 \\ {\B u}_3 \\ {\B u}_4 \ear {\small{=}} \underbrace{\lba{cccccc} {\B K}_{12} & {\B K}_{13} & {\B K}_{14} &0&0&0\\
         0&0 & 0 &  {\B K}_{23} & {\B K}_{24} & 0 \\
         0 & 0& 0 & 0 & 0 & {\B K}_{34}\\
         0 &0&0& 0 &0& 0\ear}_{=: {\B K}} \lba{c}{\B y}_1 \\ {\B y}_2 \\ {\B y}_3\\ {\B y}_4 \\ {\B y}_5 \\ {\B y}_6 \ear,
    \ee 
and $y$ is defined in \eqref{eq:exmp_meas}. 
A straightforward computation shows that $\mathcal{S}$ is not QI with respect to ${\B P}_{yu}=C_2(sI-A)^{-1}B_2$, so conversion of this output feedback controller design problem to the form $\eqref{eq:phiQ}$ is not possible.

Convex formulations of the controller design problem for this example arise when either relative measurements or network structure constraints are imposed alone, but when both properties are simultaneously present, a convex reformulation does not follow from standard methodology. The main result of this work addresses this problem. 

\section{A Relative Feedback Parameterization} \label{sec:define_relative}
{In this section, we formulate the relative measurement structure of a system as a constraint for the corresponding controller design problem.}
\textcolor{red}{
}
We begin by introducing the notion of a relative mapping and a relative LTI system. 

\begin{defn} \label{defn:rel_map}
    Let $f$ be a linear map on $\R^n$, i.e., $f(x) = Fx$ for $F \in \R^{m \times n}$. $f$ is \emph{relative} if there exists a set of vectors $\{ v_{ij}\} \subset \R^m$ such that\footnote{The choice of $v_{ij}$ in \eqref{eq:v} is non-unique, e.g., 
$
    f(x) = 3x_1 -x_2-2x_3 = 
$
 $3(x_1-x_2) + 2(x_2 - x_3) = (x_1-x_2) + 2(x_1 - x_3).$} for all $x = \lba{ccc} x_1 & \cdots & x_n \ear^{\top} \in \R^n,$
        \be \label{eq:v}
            f(x) = Fx = \sum_{1 \le i < j \le n} v_{ij} (x_i - x_j).
        \ee 
        An LTI system, ${\B H}$, 
defined by 
    \be \ba 
        \dot{\xi}(t) &= A_{H} \xi(t) + B_{H} \nu(t) \\ 
        \eta(t) & = C_{H} \xi(t) + D_{H} \nu(t) 
    \ea \ee 
    is \emph{relative} 
    if $B_{H}$ and $D_{H}$ are relative mappings. 
\end{defn}



Equipped with this terminology, 
we derive a 
parameterization of the set of LTI control policies  
    \be \label{eq:uKy}
        {\B u} = {\B K}{\B y}
    \ee 
for system \eqref{eq:P} with relative measurement structure described by $\mathcal{A}_{\rm meas.}(C_2)$. 
Because $C_2$ satisfies Assumption~\ref{assumption:C2}, we see that \eqref{eq:uKy} can be written as
    \be \label{eq:K_relationship}
        {\B u} = {\B K} C_2 {\B x} =:{\B R} {\B x},
    \ee 
where ${\B R}$ is a relative mapping. The following theorem describes the converse, characterizing conditions under which 
a control policy \eqref{eq:uKy} can be recovered from the relation \eqref{eq:K_relationship}; 
a proof of this result is provided in the Appendix.

\begin{thm} \label{prop:recover_original_K}
    Let ${\B R}$ be a relative LTI system. If $\mathcal{A}_{\rm meas.}(C_2)$ 
    corresponds to a connected graph, then there is a controller ${\B u} = {\B K}{\B y}$ for which ${\B K}C_2 = {\B R}.$ More generally, if $\mathcal{A}_{\rm meas.}(C_2)$ corresponds to a graph with $N\ge 1$ disjoint connected components, then there is a controller ${\B u} = {\B K} {\B y}$ for which ${\B K}C_2 = {\B R}$ if and only if ${\B R}$ can be written as
        \begin{equation}  \label{eq:K_graph}
                    {\B R} {\B x} = \sum_{i = 1}^N {\B R}^{(i)} {\B x}^{(i)}
        \end{equation}
            where the vector ${\B x}^{(i)}$ contains the subset of states contained in the $i^{\rm th}$ connected component of 
            $\mathcal{A}_{\rm meas}(C_2)$ and each ${\B R}^{(i)}$ is relative. We denote the structural constraint \eqref{eq:K_graph} as ${\B R} \in \mathcal{S}_{\rm rel.}(C_2).$
\end{thm}

This characterization of a relative measurement structure 
can be incorporated into the controller design problem:

\begin{cor} \label{cor:R}
    The solution ${\B K}$ to the controller design problem \eqref{eq:general} can be recovered from the solution ${\B R}$ to
    \be \label{eq:relative_formulation}\ba
        &  \inf_{\B R} ~~ \| \mathcal{F}({ \tilde{\B P}}, {\B R}) \|\\
        &~ {\rm s.t.}~{\B R} \in \mathcal{K}_{\rm stab.}(\tilde{\B P}) ~\cap~ \mathcal{S}_{\rm rel.}(C_2) 
    \ea\ee
where $ \lba{c} {\B z} \\ {\B x } \ear = \tilde{\B P} \lba{c} {\B w} \\ {\B u } \ear = \lba{cc}{\B P}_{zw} & {\B P}_{zu} \\ {\B P}_{xw} & {\B P}_{xu} \ear  \lba{c} {\B w} \\ {\B u } \ear$. 
\end{cor}


\begin{exmp} 
    Consider the system of Example~\ref{exmp:3components}, for which $\mathcal{A}_{\rm meas.}(C_2)$ had three disjoint connected components: $\{x_1, x_3\}, \{x_2, x_4\},$ and $\{x_5\}$.
By Theorem~\ref{prop:recover_original_K}, for a relative system ${\B R}$, there exists a controller ${\B K}$ satisfying ${\B R} = {\B K} C_2$ if and only if ${\B R} \in \mathcal{S}_{\rm rel.}(C_2)$, i.e.,
        \be \ba \label{eq:ex1_eq}
            {\B R} {\B x}
            & = {\B R}^{(1)} \lba{c} {\B x}_1 \\ {\B x}_3 \ear + {\B R}^{(2)}\lba{c} {\B x}_2 \\ {\B x}_4 \ear + {\B R}^{(3)} {\B x}_5,
        \ea \ee 
    with ${\B R}^{(i)}$ for $i = 1,2,3$ all relative. Since ${\B R}^{(3)}$ is a single column, it is relative only if it is zero. 
    ${\B K} = \lba{cc} {\B K}_1 & {\B K}_2 \ear$ can be recovered from ${\B R}$ as ${\B K}_1 = {\B R}^{(1)} {\small{\lba{c} 1 \\ 0 \ear}}$ and ${\B K}_2 = {\B R}^{(2)} {\small{\lba{c} -1 \\ 0 \ear}}$.
\end{exmp}

\section{A Convex Formulation of Controller Design for Relative Measurement Systems} \label{sec:convex}
{In this section, we illustrate that our relative measurement constraint can be written as a linear subspace constraint, allowing \eqref{eq:relative_formulation} to be written as a convex program.} 



\begin{prop} \label{prop:REi}
    Let $\mathcal{C}_i$, $i = 1,...,N $, denote the $N$ connected components in the graph defined by $\mathcal{A}_{\rm meas.}(C_2)$. 
    \be \label{eq:KEi}
        \mathcal{S}_{\rm rel.}(C_2) = \{{\B R}; ~  {\B R}(s)\cdot  E^{(i)} = 0, ~\forall s, ~ i = 1,...,N \}
    \ee 
where $E^{(i)}$ is the $n \times 1$ vector defined by 
    \be \label{eq:Ei}
        E^{(i)}_m = \begin{cases}
            1,~ x_m ~{\rm in~component~}\mathcal{C}_i\\
           0, ~{\rm else},
        \end{cases}
    \ee 
    with $E^{(i)}_m$ the $m^{\rm th}$ entry of the vector  $E^{(i)}.$
\end{prop}
If $\mathcal{A}_{\rm meas.}(C_2)$ corresponds to a connected graph, \eqref{eq:KEi} reduces to 
    \be \label{eq:K1}
       \mathcal{S}_{\rm rel.}(C_2) = \{ {\B R}; ~ {\B R}(s)\cdot \mathbb{1} = 0 ~ \forall s\}.
    \ee
\begin{rem}  \eqref{eq:K1} has appeared elsewhere, e.g., \cite{jensen2020gap, oral2019disorder, tegling2017coherence}. However, the extension to  relative measurement architecture that corresponds to disconnected $\mathcal{A}_{\rm meas.}(C_2)$ has not been previously considered. This extension allows the  controller ${\B K}$ solving \eqref{eq:general} to be recovered from the formulation \eqref{eq:relative_formulation} for a broader class of relative measurement structures.
\end{rem}
 
\begin{thm} \label{main_result}
The controller design problem \eqref{eq:relative_formulation} for system \eqref{eq:P} can be written equivalently as 
    \be \ba \label{eq:opt_Q}
        &\inf_{\B Q} ~~ \| {\B T}_1 - {\B T}_2 {\B Q} {\B T}_3 \|\\
        & ~ {\rm s.t.}~~{\B Q}~{\rm stable},~ {\B Q} \in \mathcal{S}_{\rm rel.}(C_2)
    \ea \ee 
where 
    \be \ba \label{eq:T}
        {\B T}_1 & = {\B P}_{zw} + {\B P}_{zu}{\B R}_{\rm nom.}(I - {\B P}_{xu} {\B R}_{\rm nom.})^{-1} {\B P}_{xw}, \\
        {\B T}_2 & = -{\B P}_{zu}(I - {\B R}_{\rm nom.}{\B P}_{xu})^{-1}, \\
        {\B T}_3 & = (I-{\B P}_{xu} {\B R}_{\rm nom.})^{-1} {\B P}_{xw},
   \ea \ee 
and the system ${\B R}_{\rm nom.} \in \mathcal{K}_{\rm stab.}{\small{\Big(\lba{cc}{\B P}_{zw} & {\B P}_{zu} \\ {\B P}_{xw} & {\B P}_{xu} \ear\Big)}} \cap \mathcal{S}_{\rm rel.}(C_2)$ is  stable. 
 The solution ${\B R}$ to \eqref{eq:relative_formulation} can be recovered from the solution ${\B Q}$ of \eqref{eq:opt_Q} as
    \be \label{eq:RfromQ}
        {\B R} = {\B R}_{\rm nom.} - h\left(h\left({\B R}_{\rm nom.}, {\B P}_{xu}\right),{\B Q}\right).
    \ee
\end{thm}

{Note that when \eqref{eq:P} is stable, ${\B R}_{\rm nom.}$ can be taken to be zero, reducing the parameters \eqref{eq:T} to ${\B T}_1 = {\B P}_{zw},$ ${\B T}_2 = -{\B P}_{zu}$ and ${\B T}_3 = {\B P}_{xw}.$
    {and reducing \eqref{eq:RfromQ} to }
    $
        {\B R} = (I + {\B Q} {\B P}_{xu})^{-1}. 
    $

The proof of Theorem~\ref{main_result} follows from the parameterization \eqref{eq:set_unstable_plant} of  stabilizing controllers and the following relation of relative measurement architectures within this parameterization.

\begin{lem} \label{thm:unstable}
    Let the systems ${\B R}$ and ${\B Q}$ satisfy $${\B R} = {\B R}_{\rm nom.} - h\left(h\left({\B R}_{\rm nom.}, {\B P}_{xu}\right),{\B Q}\right),$$ and assume that ${\B R}_{\rm nom}\in \mathcal{S}_{\rm rel.}(C_2)$. Then ${\B R}(s) \in \mathcal{S}_{\rm rel.}(C_2) $ if and only if ${\B Q}(s) \in \mathcal{S}_{\rm rel.}(C_2).$
\end{lem}

A proof of Lemma~\ref{thm:unstable} is in the Appendix.


    

\section{Incorporating Network Structure}\label{sec:incorporate_control_structure}
{
Relative measurement requirements often appear in distributed settings \cite{ahsun2006dynamics,abbasi2022decentralized,wang2018toward,jensen2020gap}, where network structural restrictions on the controller are often also present. 
We illustrate that certain network structural constraints can be incorporated into formulation \eqref{eq:opt_Q} while preserving convexity. We return to the example of Section~\ref{sec:counterexample} to illustrate this.}



\begin{exmp}\label{exmp_continued} 
By Corollary~\ref{cor:R}, the solution ${\B K}$ of
\be \ba \label{exmp_eq1}
        & \inf_{\B K} ~~ \| \mathcal{F}({\B P}; {\B K}) \|\\
        &~{\rm s.t.}~~ {\B K} \in \mathcal{K}_{\rm stab.}({\B P}) \cap \mathcal{S}_{\rm net.}\\
    \ea \ee
for system \eqref{eq:exmp_dynamics}
 can be recovered from the solution ${\B R}$ of 
    \be \ba\label{eq:exmpR}
        & \inf_{\B R} ~~ \| \mathcal{F}(\tilde{\B P}; {\B R}) \|\\
        &~{\rm s.t.}~~ {\B R} \in \mathcal{K}_{\rm stab.}(\Tilde{\B P}),\\&~~~~~~~{\B R} \in  \mathcal{S}_{\rm rel.}(C_2) = \{{\B R}; ~ {\B R}(s) \mathbb{1} = 0 \} ~ \\
       &~~~~~~~{\B R} \in \mathcal{S} := \{{\B F} C_2; {\B F} \in \mathcal{S}_{\rm net.}\}.
    \ea \ee 
A straightforward computation shows that  $\mathcal{S}$ is equivalent to the set of upper-triangular transfer matrices, 
and is QI with respect to ${\B P}_{xu}$. By Theorem~\ref{main_result}, \eqref{eq:exmpR} can be written as the convex program
    \begin{equation*}  \ba
        & \inf_{\B Q} ~~ \| {\B P}_{zw} + {\B P}_{zu} {\B Q} {\B P}_{xw} \| \\
        &~ {\rm s.t. }~~ {\B Q} ~{\rm stable}, ~{\B Q} \in \mathcal{S}_{\rm rel.}(C_2),~{\B Q} ~{\rm upper~ triangular.}
    \ea \end{equation*}
\end{exmp}
We can view this reformulation as transforming an output feedback problem to a state feedback problem with added constraints. \emph{This state feedback formulation preserves structure in the open-loop state dynamics that is lost through multiplication by $C_2$ to form the relative measurement vector.} Network control constraints may be preserved under linear fractional transformations for this \emph{structured} representation of the plant (through quadratic invariance). It is reasonable to expect this to occur in other problems, especially when the open-loop state dynamics are decoupled or dependent only on relative states, both of which occur frequently in distributed settings.
We further illustrate our result with a numerical example.

\subsection{Numerical Example}

%


We consider a distributed system composed of $n$ identical first-order, discrete-time subsystems
\be \ba \label{eq:ring_ss_dynamics}
    x_i[t+1] &=  \alpha x_i[t] + u_i[t] + w_i[t], ~~ i = 1,...,n
\ea \ee
where $\alpha\in(0,1)$ so the system is stable.
The performance output penalizes deviation from consensus and control effort:
\be \label{eq:ring_ss_regulator}
    \B{z} = (1-\gamma)\lba{c}\bar{C}_n\\0_n\ear \B{x} + \gamma\lba{c}0_n\\I_n\ear \B{u},
\ee
where $ \bar{C}_n:= I_n - \frac{1}{n}\one_n\one_n^T$ 
and  $\gamma\in[0,1]$.

{The subsystems are connected in a ring graph. Let $l(i,j)$ be the length of the shortest path between $i$ and $j$ in this graph.
The subcontroller at subsystem $i$ has access to (relative) measurements from subsystem $j$ from $m$ times steps ago only if 
$l(i,j)\leq m$. 
The corresponding set ${\mathcal{S}}_{\rm rel.}$ is $\{{\B R}; {\B R}(s) \mathbb{1} = 0\}$. To define the corresponding set $\mathcal{S}_{\rm net.},$ construct} the $m$-step adjacency matrix
$
    \A^m_{ij} = \begin{cases}
        1, & l(i,j) \leq m\\
        0, & \text{else}
    \end{cases},
$
and let ${\rm Sp}(\mathcal{A}^m)$ be the set of transfer matrices with the same sparsity pattern. 
Then,
    \be 
        \mathcal{S}_{\rm net.}:=\sum_{m=0}^n\big( {\rm Sp}(\mathcal{A}^m) \cap z^{-m}RL_{\infty}\big),
    \ee 
where  
$        z^{-m} RL_{\infty} := \{\B{H}(z) \in RL_{\infty}; ~ z^m \B{H}(z) \in RL_{\infty} \},$ and $RL_{\infty}$ denotes the set of rational, proper, transfer matrices of dimension $n \times n$. E.g., when $n=3$,
{{
\be
    \B{R}(z) = \lba{ccc} \B{r}^0_{11}&0&0\\0&\B{r}^0_{22}&0\\0&0&\B{r}^0_{33}\ear + \frac{1}{z} \lba{ccc} \B{r}^1_{11}&\B{r}^1_{12}&\B{r}^1_{13}\\\B{r}^1_{21}&\B{r}^1_{22}&\B{r}^1_{23}\\\B{r}^1_{31}&\B{r}^1_{32}&\B{r}^1_{33}\\ \ear
\ee
}}
where each $\B{r}^m_{ij}\in RL_{\infty}$.

\subsubsection{Controller Design}
It is straightforward to show that $\mathcal{S}_{net.}$ is QI with respect to ${\B P}_{xu}$. By Theorem~\ref{main_result}, we can write the  $\mathcal{H}_2$ controller design problem as a convex program:
\be\label{eq:ring_constr_opt}\ba
    &\inf_{\B Q} ~~ \| {\B P}_{zw} + {\B P}_{zu} {\B Q} {\B P}_{xw} \|_{\mathcal{H}_2}^2\\
    & ~ {\rm s.t.}~~{\B Q}\in\mathcal{S}_{\rm net.} \cap \mathcal{S}_{\rm rel.}, ~ {\B Q} ~ {\rm stable}
\ea \ee
{{To solve \eqref{eq:ring_constr_opt} numerically, we note these is no loss in restricting $\B{Q}(z)$ to have a circulant structure \cite{bamieh2002distributed}, e.g., 
for $n=3$, $\B{Q}(z) \in \mathcal{S}_{\rm net.}$ will have the form
\be\label{eq:ring_Q_form}
   \B{Q}(z) = \B{q}_0(z) I_3 
    + \frac{1}{z}
    \lba{ccc}
        0 & \B{q}_1(z) & \B{q}_2(z)\\
        \B{q}_2(z) & 0 & \B{q}_1(z)\\
        \B{q}_1(z) & \B{q}_2(z) & 0
    \ear.
\ee
This allows the objective of \eqref{eq:ring_constr_opt} to be rewritten as
$$\ba\left\|\B{P}_{zw}+\B{P}_{zu}\B{Q}\B{P}_{xw}\right\|_{\mathcal{H}_2}^2 
    &= n\left\|\B{P}_{zw}e_1+\B{P}_{zu}\B{P}_{zw}(\B{Q}e_1)\right\|_{\mathcal{H}_2}^2\ea$$
To impose the relative constraint $\B{Q} \in \mathcal{S}_{\rm rel.}$ we write $\B{q}_0$ in terms of the other free parameters, e.g., for $n=3$, $\B{q}_0 = -\frac{1}{z}\B{q}_1-\frac{1}{z}\B{q}_2$.
Defining $\B{q} = \lba{ccc} \B{q}_1 & \dots & \B{q}_{n-1} \ear^T$ and $\B{W}$ as the transfer matrix that satisfies $\B{Q}e_1 = \B{W}\mathbf{q}$, we write \eqref{eq:ring_constr_opt} as an unconstrained model-matching problem \cite{francis1987course}: 
    \be  \label{eq:modelmatching}
       J_n =  \inf_{{\B q}~{\rm stable}}~ n\| {\B P}_{zw} e_1 + {\B P}_{zu} {\B P}_{xw} {\B W}{\B q}\|_{\mathcal{H}_2}^2.
    \ee 
The optimal norm $J_n$ is computed for various values of $n$ with $\gamma=0.5$ and $\alpha=0.6$ (See Figure~\ref{fig:ctrl_performance}). 
    \setlength{\belowcaptionskip}{-15pt}
\begin{figure} 
    \centering
\includegraphics[width=.9\linewidth]{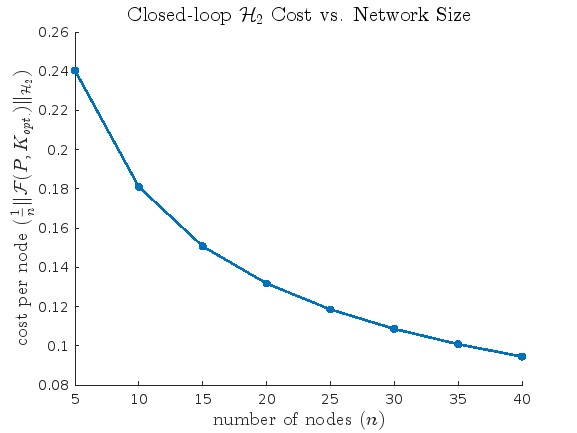}
    \caption{Cost (with $\gamma=0.5$, $\alpha=0.6$) per node ($\frac{1}{n}||\mathcal{F}({\B P}, {\B K}||$) for the $\mathcal{H}_2$ optimal controller solving \eqref{eq:modelmatching}.}
    \label{fig:ctrl_performance}
\end{figure}

\section{Conclusion}
We provided a characterization of relative measurement structures, which are common in distributed systems, as a constraint. This allowed the controller design problem to be written as an equivalent convex program, even when certain additional network structural constraints were present. Ongoing work will extend to allow for more general relative measurements, e.g. differences of outputs rather than differences of states. More generally, we aim to utilize this formulation as a step toward understanding fundamental properties of relative measurement systems such as best achievable performance.

\bibliography{bibliography}
\bibliographystyle{ieeetr}

\section*{Appendix}

\subsection*{Proof of Theorem~\ref{prop:recover_original_K}}
We utilize the following lemmas, the proofs of which are ommitted for space.
\begin{lem}\label{app_lem1}
    Any relative linear map $f(x) = Fx$ on $\mathbb{R}^n$ can be written as 
$        
            f(x) = \sum_{i = 1}^{n-1} g_i (x_i - x_{i+1} ),
        $
    for some set of vectors $\{g_i\}.$
\end{lem}
\begin{lem} \label{app_lem2}
If 
$\mathcal{A}_{\rm meas.}(C_2)$ computed from \eqref{eq:Ameas}  corresponds to a connected graph, then there is an invertible matrix $T$ for which 
\small{\be \label{eq:M} (T C_2) x = \underbrace{\lba{ccccc} 1 & -1 \\ & 1 & -1 \\ & & \ddots & \ddots \\ & & & 1 & -1 \ear}_{=: M} \lba{c} x_1 \\ x_2 \\ \vdots \\ x_n \ear =  \lba{c} x_1 - x_2 \\ x_2 - x_3 \\ \vdots \\ x_{n-1} - x_n \ear.\ee}
\end{lem}

\begin{lem} \label{lem:markov}
    Consider two LTI systems ${\B K}$ and ${\B R}$ with Markov parameters $\{D_K, C_KB_K, C_K A_K B_K,...\}$ and $\{D_R, C_RB_R, C_R A_R B_R,...\}$, respectively. For a static matrix $C_2,$ the relation 
        $
            {\B K} C_2 = {\B R}
        $
    holds if and only if the following matrix equalities hold
    \vspace{-1mm}
        \be \ba
            &D_K C_2 = D_R, 
            & C_K A_K^i B_K C_2 = C_R A_R^i B_R, ~ i \ge 0.
        \ea \ee 
\end{lem}
First, note that there is a solution $G$ to the matrix equation 
\be \label{eq:GF}
    GM = F,
\ee 
where $M$ is defined in \eqref{eq:M}
{if and only if} $F$ is relative. For one direction of this implication, note that if $GM = F$, then $0 = GM \mathbb{1} = F \mathbb{1}$ so that $F$ is relative. Conversely, use Lemma 2 to write $GMx = Fx = \sum_{i = 1}^{n-1} g_i (x_i - x_{i+1} )$ for some vectors $\{g_i\}$ and any $x \in \mathbb{R}^n$ - then $G =\lba{cccc} g_1 & g_2 & \cdots & g_{n-1} \ear$ solves \eqref{eq:GF}.
Then, by Lemma 4, a solution $\tilde{\B K}$ to 
    \be 
        \tilde{\B {K}} M = {\B R}
    \ee
exists if and only if ${\B R}$ is relative. By Lemma 3, since ${\B K} C_2 = {\B R}$ can always be written as $ {\B K} T^{-1} M = {\B R}$ for some $T,$ a solution ${\B K}$ to ${\B K} C_2 = {\B R}$ exists if and only if ${\B R}$ is relative. This completes the proof of Theorem~\ref{prop:recover_original_K} in the case that $\mathcal{A}_{\rm meas.}$ is connected.

When $A_{\rm meas.}$ corresponds to a graph with $N >1 $ connected components, and ${\B R}$ satisfies \eqref{eq:K_graph}, we search for ${\B K}$ satisfying
\vspace{-2mm}
    \be \label{eq:app_Ncomponents}
        {\B K } C_2 {\B x} = \sum_{i = 1}^N {\B R}^{(i)} {\B x}^{(i)}
    \ee 
for all ${\B x}.$ The left hand side of \eqref{eq:app_Ncomponents} is 
{\setlength{\arraycolsep}{2pt}
{\small{    \begin{equation} 
        \lba{cccc}{\B K}_1  & \cdots & {\B K}_N \ear  \lba{cccc} M^{(1)} \\    & \ddots\\  & & M^{(N)} \ear \lba{c} {\B x}^{(1)}  \\ \vdots \\ {\B x}^{(N)} \ear\end{equation}}}}
where each $M^{(i)}$ is of the same form as $M$ \eqref{eq:M}, and ${\B K}$ is block partitioned accordingly. This problem reduces to solving $N$ problems of the form ${\B K}_i M^{(i)} = R^{(i)}$, which follows from  the connected graph setting. 
\color{black}

\subsection*{Proof of Lemma~\ref{thm:unstable}}

    Rearranging \eqref{eq:set_unstable_plant},  ${\B Q}$ can be recovered from ${\B R}$ as 
        $
            {\B Q} = (I - ({\B R}_{\rm nom.} - {\B R}) {\B N})^{-1} ({\B R}_{\rm nom.} - {\B R}),
        $
    where ${\B N} = {\B P}_{xu} (I - {\B R}_{\rm nom.} {\B P}_{xu})^{-1}.$ Then, since ${\B R}_{\rm nom.} E^{(i)} = 0, $ ${\B R} E^{(i)} = 0 ~ \Rightarrow ~Q E^{(i)} = 0.$ Conversely, if ${\B Q} E^{(i)} = 0,$ then 
        \be \ba
            {\B R} E^{(i)}& =\left( {\B R}_{\rm nom.} - h\left({\B V},{\B Q}\right)\right)  E^{(i)}\\
            & = -(I - {\B Q}{\B V})^{-1}{\B Q} E^{(i)} = 0, 
        \ea \ee 
    where ${\B V} = h\left({\B R}_{\rm nom.}, {\B P}_{xu}\right).$ \hfill $\blacksquare$


\end{document}